\documentclass[fleqn,usenatbib]{mnras}
\usepackage{newtxtext,newtxmath}

\usepackage{CJK}
\usepackage{bm}
\usepackage{units}
\usepackage{amsmath}
\usepackage{float}
\usepackage{pgfplots}
\usepackage{multirow}
\usepackage{enumitem}
\pgfplotsset{compat=1.15}

\usepackage{xcolor}
\definecolor{Garnet}{rgb}{0.6, 0.1, 0.2}
\newcommand{\textmod}[1]{#1}

\renewcommand{\theenumi}{\arabic{enumi}}

\title[Heliospheric $1/f$ spectrum: superposition]{$1/f$ noise in synthetic and solar wind data: superposition principles}

\author[J. Wang et al.]{Jiaming Wang (王嘉明),$^{1}$\thanks{E-mail: jmwang@udel.edu} Francesco Pecora,$^{1}$
Rohit Chhiber,$^{1, 2}$
Rayta A.\ Pradata,$^{1}$
Subash Adhikari$,^{1}$
\newauthor
William H.\ Matthaeus$^{1}$
\\
$^{1}$Department of Physics and Astronomy, University of Delaware, Newark, DE 19716, USA\\
$^{2}$Heliophysics Science Division, NASA Goddard Space Flight Center, Greenbelt, MD 20771, USA \\
}

\date{Accepted XXX. Received YYY; in original form ZZZ}

\pubyear{XXX}

\begin{document}
\begin{CJK*}{UTF8}{gbsn}


\label{firstpage}
\pagerange{\pageref{firstpage}--\pageref{lastpage}}

\maketitle

\end{CJK*}

\begin{abstract}

The interplanetary magnetic field exhibits
a distinctive $1/f$ spectral density from frequencies of around $\unit[10^{-6}]{Hz}$ to around $\unit[10^{-4}]{Hz}$, ranging from harmonics of the solar rotation to the reciprocal of the turbulence correlation time in the spacecraft frame. Various theories have been proposed to explain its origin, typically invoking either processes in the lower corona or in the solar interior, or local interplanetary dynamics. Here, we investigate the {\it superposition principle} that underlies explanations of the solar/coronal types, which in principle can generate the full observed range of $1/f$ noise. Using synthetic time series with scale-invariant or lognormal distributions of correlation times, we examine the efficacy of several superposition approaches in generating a $1/f$ regime. The persistence of $1/f$ spectrum is further illustrated with decade-long {\it in situ} magnetic field measurements from the ACE spacecraft. Together, these results help explain the ubiquity of $1/f$ noise under the unavoidable superposition inherent in long-duration heliospheric data.

\end{abstract}


\section{Introduction}
\label{sec:intro}


The presence of ``$1/f$ noise'', or {\it flicker noise}, in the interplanetary medium is a familiar observation, while at the same time, its origin remains elusive. Generally associated with a power-law spectral density in the frequency $f$ domain, this phenomenon often spans more than a decade in scale. At 1 au, $1/f$ signals have been observed in fluctuations of the magnetic field~\citep{Matthaeus86, Feynman94} and in the plasma density~\citep{Matthaeus07}. $1/f$ noise in magnetic field persists into the inner heliosphere as indicated by measurements from MESSENGER and the Parker Solar Probe~\citep[PSP;][]{Davis23, Huang23, Pradata25, Huang25}. Beyond {\it in situ} measurements, $1/f$ noise has been identified in remote sensing data using line of sight photospheric magnetic field~\citep{Nakagawa74, Matthaeus08} and inferred in channel-integrated SOHO/UVCS data~\citep{Bemporad08}. 
NASA's Parker Solar Probe (PSP) Science and Technology Definition team identified the potential discovery of the origin of the interplanetary $1/f$ signal as a major goal for the PSP mission~\citep{plusreport08}. 

Discussions of the origin of $1/f$ noise in the interplanetary medium may be roughly divided into two classes: those attributing it to local {\it in situ} dynamical processes~\textmod{\citep[see, e.g.,][]{Velli89, Verdini12, Matteini18, Chandran18, Huang23}}, and those linking it to remote activities, perhaps in the sub-Alfv\'enic corona or even within the solar interior beneath the photosphere\textmod{\citep[see, e.g., ][]{Matthaeus86, Dmitruk14, Wang24_1overf}}. This paper sets its context in the latter category, hereafter referred to as the {\it solar origin} for brevity.

Beyond the solar wind, a great variety of other systems also display $1/f$ signature~\citep[see, e.g., review by][]{Dutta81, Wang24_1overf}, rendering it a subject of considerable study across disciplines. Of particular interest for broad applicability are the suggestions of generic principles 
that can be responsible for producing a $1/f$ signal but do not require specific dynamical mechanisms. Generic principles include the superposition
principle that will be the main subject here as initially described by \citet{vandeZiel50} and \citet{Machlup81}, as well as ideas based on self-organized criticality phenomenon, including the ``sandpile'' models 
\citep[see, e.g., ][]{Bak87, Hwa92, Consolini15, Korzeniowska23}. There has been substantial discussion in the literature regarding possible relationships or distinctions between these two general classes~\citep[see, e.g.,][]{Bruno13, Lamarre25}. Other models for generating scale-invariant signals either in time (to obtain $1/f$ in frequency) or in space ($1/k$ in wavenumber $k$) have been discussed, often in the context of self-organization
or inverse cascade. These phenomena are typically slow, requiring many nominal nonlinear times~\citep{Brandenburg01, Dmitruk07, Dmitruk11}. 
There are also models that invoke local processes or instabilities, and some balance between spectral transfer and expansion~\citep{Velli89, Huang23, Huang25}. Such local models may be self-consistent, but are intrinsically limited in their range of influence. That is, they may not be able to explain very low-frequency spectra due to limited solar wind propagation time~\citep{Zhou90, Chhiber2018thesis}.

The main purpose here is to examine \textmod{the feasibility of generating $1/f$ signals through superposition carried out explicitly in the time domain.} The widely quoted 
superposition ideas of \citet{vandeZiel50} and \citet{Machlup81}, 
along with enabling frequent appearance of lognormal distributions of relevant physical quantities~\citep{Montroll82}, are couched in applied mathematical terms. Usual descriptions involve superposed exponential correlation functions and associated spectra, as briefly outlined in the following section. However, \textmod{how such a superposition procedure is physically achieved} is rarely explicated in any detail, even if it appears to be reasonable on an intuitive level. Therefore, this paper focuses on examination of several versions of \textmod{time-domain signal superposition using synthetic time series endowed with scale-invariant time-scales. In this way, our simplified system touches base with the pristine solar wind, which admits several kinds of scale invariance. The coronal solar wind and the solar interior dynamo may also share these properties, so the superposition procedure could plausibly operate in those environments as well.}

In the solar wind and coronal context, superposition might involve successive magnetic flux tube merging that creates a lognormal distribution of correlation scales~\citep{Matthaeus86}. 
Recently, lognormal correlation scales have been observed in the solar photosphere and the low corona~\citep{Sharma23, Chhiber25_vKH}. The requisite large scales as indicated by the low-frequency extent of the observed interplanetary $1/f$ band may also be linked to processes in the solar dynamo, as motivated by dynamo simulations and experiments~\citep{Ponty04, Dmitruk14}. Or they may be linked to inverse cascade systems known to produce time-scales much greater than the band-limited scales of the system~\citep{Dmitruk07, Dmitruk11}. In this regard, the limitation on available dynamical time-scales could be alleviated, given that the relevant mechanisms operate in sub-Alfv\'enic coronal regions, where a parcel of plasma may reside for many nonlinear times, rather than in the super-Alfv\'enic wind.

Here, we do not attempt \textmod{a} detailed explanation of the exact mechanism that creates the lognormal distribution underlying the superposition procedure, but look at different ways signals may be superposed and their effectiveness in generating 1/f noise, without prejudice concerning the location at which this occurs. \textmod{We will show with synthetic data that superposition in the time domain, in several different interpretations, is a feasible route to the measured 1/f spectrum. We will then show using in situ measurements that the application of the same superposition operations preserves the 1/f spectrum.} The paper is organized as follows. In the next section, we describe the fundamental superposition principle. In Section~\ref{sec:synthetic}, several versions of superposition are described and analysed in the context of synthetic data sets, with details of their construction provided in the Appendix. In Section~\ref{sec:observation}, these superposition versions are studied employing a 12-yr NASA's Advanced Composition Explorer (ACE) data set at 1 au. Section~\ref{sec:conclusion} summarizes and discusses the results.

\section{Overview of the superposition principle}
\label{sec:principle}

It has been demonstrated by~\citet{Machlup81} that the superposition of \textmod{the so-called Ornstein--Uhlenbeck processes -- characterized by} exponentially decaying autocorrelations of the form $e^{-\tau/\tau_c}$ where $\tau_c$ is the characteristic correlation time -- results in a $1/f$ power spectrum, provided that the ensemble of $\tau_c$ follows a scale-invariant distribution. \textmod{The Ornstein--Uhlenbeck process is the unique stationary Gaussian process in which the future evolution depends solely on the present state and not on the past history. It therefore arises naturally as a simple model for stochastic systems.} The autocorrelation function of a zero-mean time series $x(t)$ can be defined as $R(\tau) = \langle x(t) x(t+\tau) \rangle$, where $\tau$ represents the time lag. Then, the Fourier transform of a single exponential autocorrelation function $R(\tau_c, \tau) \propto e^{-\tau/\tau_c}$ in angular frequency $(\omega = 2\pi f)$ domain yields a spectrum of Lorentzian form:
\begin{equation}
    S(\tau_c, \omega) = \int_{-\infty}^\infty e^{-i\omega \tau} R(\tau_c, \tau) d\tau \propto \frac{2 \langle x^2 \rangle \tau_c}{1+\tau_c^2 \omega^2},
\label{eq:Machlup1}
\end{equation}
where $R(\tau)$ is an even function symmetric about the origin, assuming the homogeneity of the data. Hereafter, \textmod{for simplicity,} we will assume 
a unit variance $\langle x^2 \rangle = 1$ unless otherwise specified.\footnote{\textmod{A scale-invariant distribution of variance preserves the analytical $1/\omega$ behaviour (as in Eq.~\ref{eq:Machlup2}). Additionally, if the variance scales with correlation time as $\langle x^2 \rangle \propto \tau_c^\beta$ with $|\beta| < 1$, the resulting superposed spectrum as in Eq.~\ref{eq:Machlup2} follows the form of $\omega^{-(1+\beta)}$. The exponent $\beta$ may vary between systems, and a detailed characterization is beyond the scope of this paper.}}
When multiple spectra of equal integrated power are superposed under the assumption that $\tau_c$ follows an inverse (scale-invariant) distribution $\rho(\tau_c) d\tau_c \propto d\tau_c/\tau_c$ within some specified domain $[\tau_1, \tau_2]$, the resulting overall spectrum is 
\begin{equation}
    \langle S (\omega) \rangle = \int_{\tau_1}^{\tau_2} S(\tau_c, \omega) \rho(\tau_c) d\tau_c \propto \frac{\tan^{-1}(\tau_c \omega)}{\omega}\Big|_{\tau_c = \tau_1}^{\tau_2}.
\label{eq:Machlup2}
\end{equation}
Assuming $\tau_1 \ll \tau_2$, and within the frequency range where $1/\tau_2 \ll \omega \ll 1/\tau_1$, Eq.~\ref{eq:Machlup2} leads to
\begin{equation}
    \langle S (\omega) \rangle \propto \frac{\frac{\pi}{2} + \mathcal{O}(\frac{1}{\tau_2 \omega})- \mathcal{O}(\tau_1 \omega)}{\omega}.
\label{eq:Machlup3}
\end{equation}
Thus under zeroth-order approximation, $\langle S \rangle \propto 1/\omega$ where $1/\tau_2 \ll \omega \ll 1/\tau_1$.

\textmod{The model considered up to this point} is often interpreted to \textmod{mean} that scale-invariant phenomena arise from processes lacking a characteristic scale \textmod{-- the frequency range of scale-invariance is set by the span of system correlation time-scales.}

\citet{Montroll82} later pointed out that an inverse distribution is not normalizable without imposing lower and upper bounds, which may be unrealistic in many natural systems lacking sharp boundaries, including the solar wind. A lognormal distribution with sufficiently large variance has been proposed as a suitable substitution. Adopting the condition in \cite{Wang24_1overf} that a lognormal random variable $x\sim \text{logNormal}(\mu,\sigma^2)$ within the domain such that $|\ln{(x/\overline{x})}| \leq \sqrt{2\theta \sigma^2}$, where $\overline{x} \equiv e^{\mu}$, is scale-invariant within a factor of tolerable deviation $\theta$, then the boundaries of the scale-invariant portion are $\exp(\mu \pm \sqrt{2\theta \sigma^2})$. In this simplistic perspective, the $1/f$ regime spans for approximately a factor of $\exp(2\sqrt{2\theta\sigma^2})$, independent of $\mu$. The correlation time at 1 au is lognormal with the variance of the underlying normal distribution being around unity (see Section~\ref{sec:observation}, also see \citet{Ruiz14}). A $\theta$ value of 0.5 would then yield nearly three-quarters of a decade of $1/f$ range in the solar wind spectrum.\footnote{The observed interplanetary $1/f$ band extends over more than a decade down to $\unit[10^{-6}$]{Hz}, which implies the presence of correlations on time-scales substantially longer than the local turbulence correlation time typically inferred from the $1/e$ method in the literature and considered here (see Section~\ref{sec:observation}).}
 

While Machlup's $1/f$ model is broadly relevant, 
the solar wind turbulence spectra are often not Lorentzian as in Eq.~\ref{eq:Machlup1}. For example, isotropic and incompressible hydrodynamic turbulence in the inertial range acquires the Kolmogorov $k^{-5/3}$ power-law scaling, which is frequently observed as $f^{-5/3}$ in time domain in the interplanetary spectrum~\citep[see, e.g., review by][]{Bruno13}. Towards lower heliocentric distances with increasing magnetic field strength, inertial range spectrum shifts to a near $f^{-3/2}$ scaling~\citep[see, e.g., recent work by][]{Chen20, Davis23} consistent with \textmod{theories} of magnetohydrodynamic (MHD) turbulence~\textmod{\citep{Biskamp03, Beresnyak19, Schekochihin22}}. \textmod{\citet{Wang24_1overf} present} a generalized version of the aforementioned superposition principle, in which the $1/f$ noise emerges
from superposing power laws of arbitrary index less than $-1$, thereby justifying the applicability of the superposition principle in the solar wind. In the next section, we generate time series with $f^{-5/3}$ power laws to test the robustness of the superposition principle.

\section{Superposition with Synthetic Data}
\label{sec:synthetic}

The \textmod{superposition invoked} in the \textmod{idealized} \cite{Machlup81} \textmod{construction} refers specifically to taking the ensemble mean of the spectra, or equivalently, the correlations. \textmod{However, this procedure does not map onto any literal physical processes in the solar wind: the plasma does not superpose correlation functions, nor do spacecraft measurements directly implement such an operation. Nevertheless, in situ solar wind measurements (e.g. the magnetic field time series) are unavoidably the outcomes of some form of superposition at the level of signals, either in generation or in transport.} Long-duration measurements necessarily \textmod{aggregate} plasma patches originating from a range of solar latitudes and longitudes, \textmod{which are} then deflected and \textmod{intermixed en route to 1 au}~\citep[see, e.g., ][]{Levine77, Servidio14, Chhiber21}. Consequently, samples widely separated in time and recorded by a single spacecraft are essentially uncorrelated \textmod{in detail}, even if they broadly share statistical \textmod{properties inherited from} their common source in the solar interior or dynamo. Additionally, wind streams from a fixed location on the solar surface can still \textmod{evolve} over time-scales of several days, partly \textmod{due to} the \textmod{continual reconfiguration of the photospheric magnetic carpet} on around $ 40$-h time-scales~\citep{Priest02}. \textmod{Given} the need of month- to year-long intervals to \textmod{resolve} the full bandwidth of heliospheric $1/f$ noise (as discussed in Section~\ref{sec:intro}), \textmod{it becomes essential to assess whether distinct forms of superposition can lead to the same $1/f$ scaling.}

\textmod{In this section, we explore} four methods of superposition using synthetic time series: 
\begin{enumerate}[nosep]
    \item averaging the correlation functions,
    \item averaging the time series,
    \item concatenating the collection of time series, and
    \item truncating each time series to a random (uniformly distributed) length between one and one-twentieth of the interval duration, then concatenating the truncated segments.
\end{enumerate} 
Note that Method 1 aligns with Machlup's procedure. Different methods may capture different aspects of solar wind dynamics, depending on the nature of the underlying variability that generates 
approximate scale-invariance or lognormality.

\subsection{Generation of synthetic time series}

We create 500 sets of random time series, each having $N = 1440$ data points to mimic a 1-d data set with 1-min resolution, and each with a mean of zero and a variance of unity. Each time series is specified with a characteristic correlation time $\tau_c = 1/ f_\mathrm{break}$, where $f_\mathrm{break}$ is an onset frequency below which the spectrum is flat and above which the spectrum has a $f^{-5/3}$ scaling. The collection of $\tau_c$ values follows a scale-invariant (or inverse) distribution of the form 
\begin{equation}
    \rho(\tau_c) d\tau_c = \frac{d\tau_c}{\tau_c \ln{(\tau_2/\tau_1)}},
\end{equation}
where $\tau_c$ spans from $\tau_1$ to $\tau_2$. From Eq.~\ref{eq:Machlup3}, $1/f$ band in the superposed data is expected to span from several factors above $1/2\pi \tau_2$ to several factors below $1/2\pi \tau_1$ in frequency space. We also analyse time series generated from the overlapping lognormal distribution, with parameters $\mu \approx [\ln(\tau_1\tau_2)]/2$ and $\sigma^2 \approx [\ln (\tau_2/\tau_1)]^2/8\theta$.

A time series $x_j$, $j = 0, \cdots, N-1$, is generated using the familiar procedure of (1) specifying a power spectrum $S_m$; 
(2) using the power spectrum to determine the Fourier amplitudes $X_m$ of the target data and then randomizing its phases; and (3) arriving at the time series through inverse Fourier transform. The procedure is described in detail in Appendix~\ref{app}.

Since the generation and analysis of the synthetic data are independent of the value of the time resolution $\Delta t$, we normalize all variables by $\Delta t$. This yields dimensionless variables (represented by overline notation) $\overline{\Delta t} = 1$, $\overline{t} = t/\Delta t$, $\overline{f} = f \Delta t$, $\overline{S}_m = S_m / \Delta t$, and $\overline{X}_m = X_m/\Delta t$. 
For simplicity, we carry out this normalization and omit the overline notation from this point onward unless stated otherwise.

We present results for three distributions of $\tau_c$: (i) inverse distribution with $\tau_1, \tau_2 = 3, 150$ to show an extended $1/f$ range for clear comparison of four methods of superposition; (ii) inverse distribution with $\tau_1, \tau_2 = 20, 150$ to mirror 1 au correlation times in minutes (see fig.~4 of \citet{Ruiz14}); and (iii) lognormal distribution with $\mu, \sigma^2 = 4, 1$ with a scale-invariant portion that overlaps approximately with distribution (ii) employing $\theta = 0.5$. For case (i), the minimum break frequency is $1/\tau_2 \approx 7 \times 10^{-3}$, which is sufficiently greater than the frequency resolution of $\Delta f = 1/N \approx 7 \times 10^{-4}$, so that the smallest frequency break can be resolved.

\subsection{Analysis of synthetic time series}
\label{sec:synthetic_analysis}

Having obtained the synthetic time series, the normalized two-point correlation function, defined as 
\begin{equation}
    R_j = \frac{\langle x_i x_{i+j} \rangle - \langle x_i \rangle \langle x_{i+j} \rangle}{\langle x_i^2 \rangle - \langle x_i \rangle^2},
\label{eq:R}
\end{equation}
is computed using the Blackman-Tukey algorithm~\citep{Blackman58, Matthaeus82-convergence}. Note that the normalization as in the denominator is redundant for data with unit variance. The operation $\langle \cdots \rangle$ denotes ensemble expectation, which is implemented in practice as averaging over the data domain indexed by $i$~\citep[see details in][]{Wang24_density}. The lag index $j$ ranges from 0 to 30 per cent of the duration of $x_j$. For individual time series, this covers nearly three times the maximum correlation time while maintaining an adequate statistical weight for each correlation value. For stationary data, the correlation value is independent of shifts in the index $i$, that is, independent of the origin of time~\citep{Matthaeus82-convergence}.

The power spectral density $S(f)$ is defined as the Fourier transform of the symmetrized correlation function. Operationally, since the correlation function is obtained on a discrete grid, the spectrum is discretized as
\begin{equation}
    S_m = \sum_{j=0}^{2M-1} e^{-i2\pi m j /2M} R_j,
\label{eq:psd_dsc}
\end{equation}
where $M = 432$ (30 per cent of the 1440 data points), and $R_{2M-j} = R_j$ for $j = M+1, \cdots, 2M-1$. It can be shown that Eq.~\ref{eq:psd_dsc} is equivalent to 
\begin{equation}
    S_m = \sum_{j=-M}^{M-1} e^{-i 2\pi mj/2M} R_j,
\label{eq:psd_dsc_2}
\end{equation}
where $R_j = R_{-j}$. We choose Eq.~\ref{eq:psd_dsc} for convenience in numerical implementation.

To prevent $R_j$ from being artificially treated as periodic over the finite domain $[-N, N]$, we first apply a 10 per cent cosine-taper window, as described in \citet{Wang24_density}. We then extend the correlation function by zero-padding its tail, imitating an infinite domain~\citep{Matthaeus82-convergence}. The tapering and zero-padding procedures reduce spectral noise at high frequencies approaching the maximally resolved (Nyquist) frequency. 
This procedure 
has negligible impact on the shape of the spectrum within the range of interest (not shown).

The spectra produced by Methods 2, 3, and 4 necessarily show substantial fluctuations. To examine the spectral shape underlying these statistical fluctuations, we apply a moving-average filter iteratively 10 times. At each iteration, the value of $S_m$ is updated to $0.5 S_m + 0.25 (S_{m-1} + S_{m+1})$~\citep{Matthaeus82-convergence}.

\subsection{Results}

\begin{figure}
\centering
    \includegraphics[angle=0,width=\columnwidth]{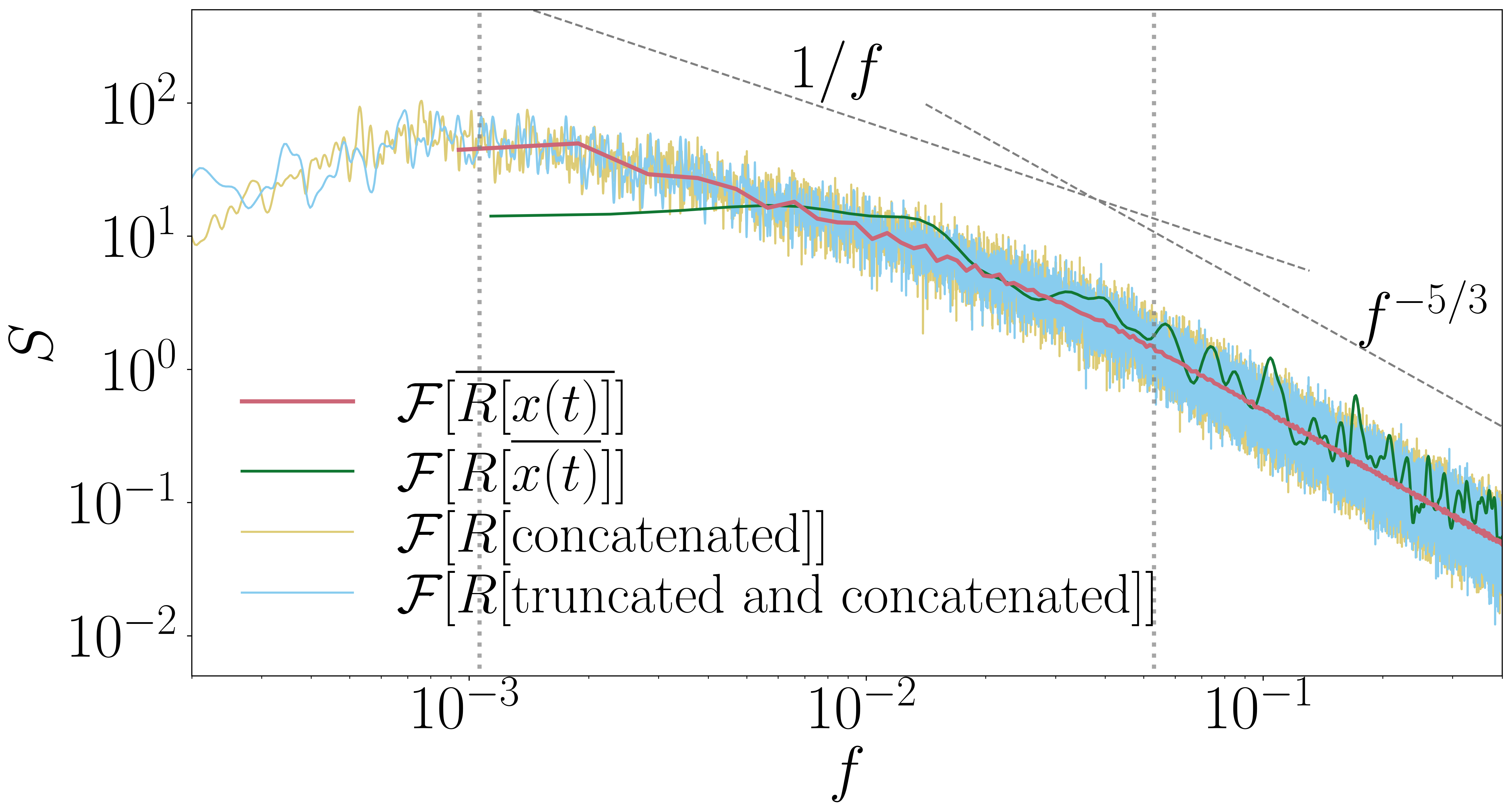}
    \includegraphics[angle=0,width=\columnwidth]{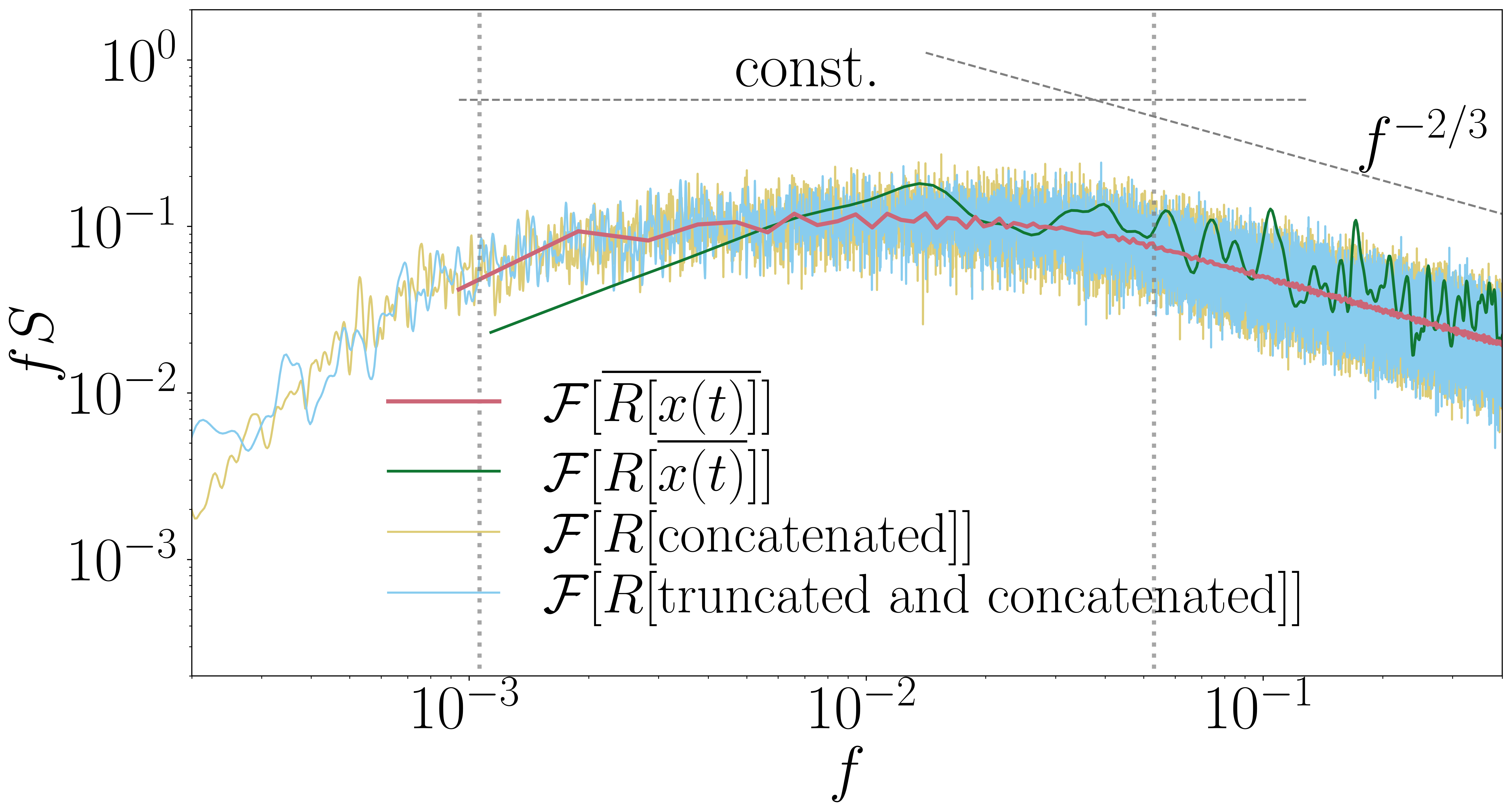}
    \caption{Top panel: power spectra resulting from scale-invariant correlation times with $\tau_1, \tau_2 = 3, 150$, obtained through superposition methods 1 (in red), 2 (in green), 3 (in yellow), and 4 (in blue). Bottom panel: compensated spectra corresponding to those in the top panel. Dotted vertical lines mark frequencies $1/2\pi \tau$ associated with the minimum and maximum correlation times, respectively. Dashed grey lines are guiding power laws with index $-1$ and $-5/3$, respectively.}
\label{fig:Ssynthetic1}
\end{figure}

For a correlation time distribution of type (i) with $\tau_1, \tau_2 = 3, 150$, the top panel of Fig.~\ref{fig:Ssynthetic1} illustrates the spectra resulting from the different superposition methods. Method 1, averaging the correlation functions (labelled as $\mathcal{F}[\overline{R[x(t)]}]$), has spectrum retaining the $-5/3$ power-law index at frequencies $f \gg 1/2\pi \tau_1$, and gradually transitions to a $-1$ index at lower frequencies. At even lower frequencies where $f \sim 1/2\pi \tau_2$, the spectrum flattens before being affected by the limited frequency resolution. The frequencies at $1/2\pi \tau_1$ and $1/2\pi \tau_2$ are indicated by dotted vertical lines.

In contrast, Method 2 -- averaging the time series (labelled as $\mathcal{F}[R[\overline{x(t)}]]$) -- exhibits no discernible $1/f$ band, showing only a weak spectral shallowing before the transition to the uncorrelated flat band. Agreement between Methods 1 and 2 is generally not expected, as it would require the ensemble of time series $\{x(t)\}$ to satisfy $\overline{\langle x(t) x(t+\tau)\rangle} \sim \langle \overline{x}(t) \overline{x}(t+\tau) \rangle$, where $\langle \cdots \rangle$ denotes an ensemble (inter-series) average and $\overline{\cdots}$ denotes a temporal (intra-series) average. An equivalence would imply the convergence of cross-correlation between data sets to autocorrelation and, ultimately, self-averaging of the ensemble. Or it would simply imply an absence of cross-correlation. Neither properties is enforced here.

The superposed spectra resulting from Method 3 (concatenating the time series) and Method 4 (truncating each time series before concatenation) are also shown in the top panel of Fig.~\ref{fig:Ssynthetic1}. Despite enhanced fluctuations from long sample duration, both methods produce spectral shapes consistent with those from Method 1. Comparing Methods 3 and 4 shows that arbitrary truncation has minimal impact on the superposed spectrum, in both the $1/f$ and the $f^{-5/3}$ bands where $f \gg 1/2\pi \tau_2$, even when some segment durations are shorter than their characteristic correlation times. Additionally, varying the lower bound of the possible truncation range from 20 per cent to 1 per cent of the 1440 data points results in no detectable change in the spectral shape (not shown). This finding is particularly relevant to spacecraft measurements, suggesting that $1/f$ signals can be recovered even when data gaps or irregular sampling renders some intervals highly fragmented. To highlight the emergence of a $1/f$ regime, we show the corresponding compensated spectra from the four methods in the bottom panel of Fig.~\ref{fig:Ssynthetic1}. 

\begin{figure}
\centering
    \includegraphics[angle=0,width=\columnwidth]{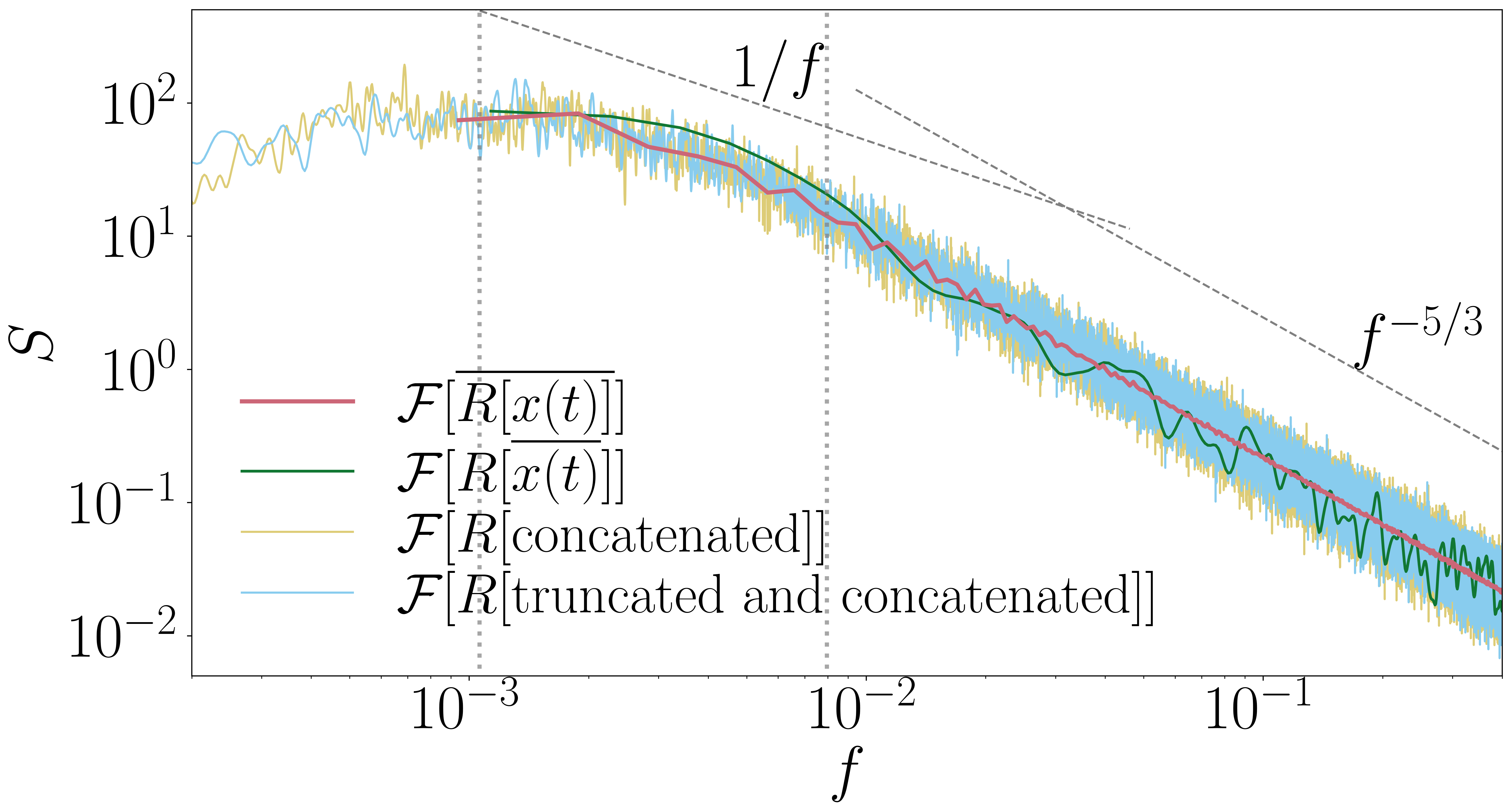}
    \includegraphics[angle=0,width=\columnwidth]{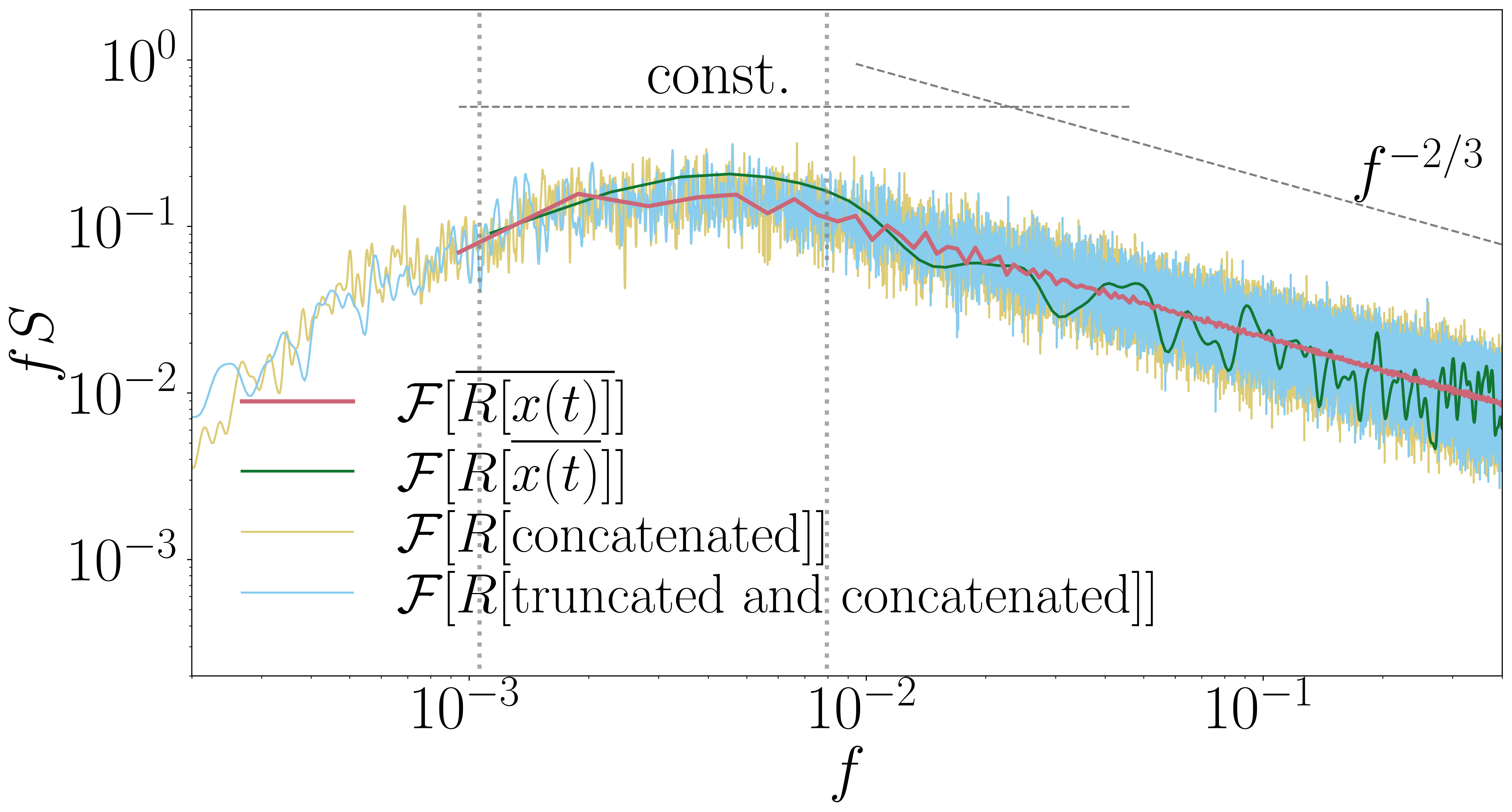}
    \caption{Top panel: power spectra resulting from scale-invariant correlation times with $\tau_1, \tau_2 = 20, 150$, obtained through superposition methods 1 (in red), 2 (in green), 3 (in yellow), and 4 (in blue). Bottom panel: compensated spectra corresponding to those in the top panel. Dotted vertical lines mark frequencies $1/2\pi \tau$ associated with the minimum and maximum correlation times, respectively. Dashed grey lines are guiding power laws with index $-1$ and $-5/3$, respectively.}
\label{fig:Ssynthetic2}
\end{figure}

\begin{figure}
\centering
    \includegraphics[angle=0,width=\columnwidth]{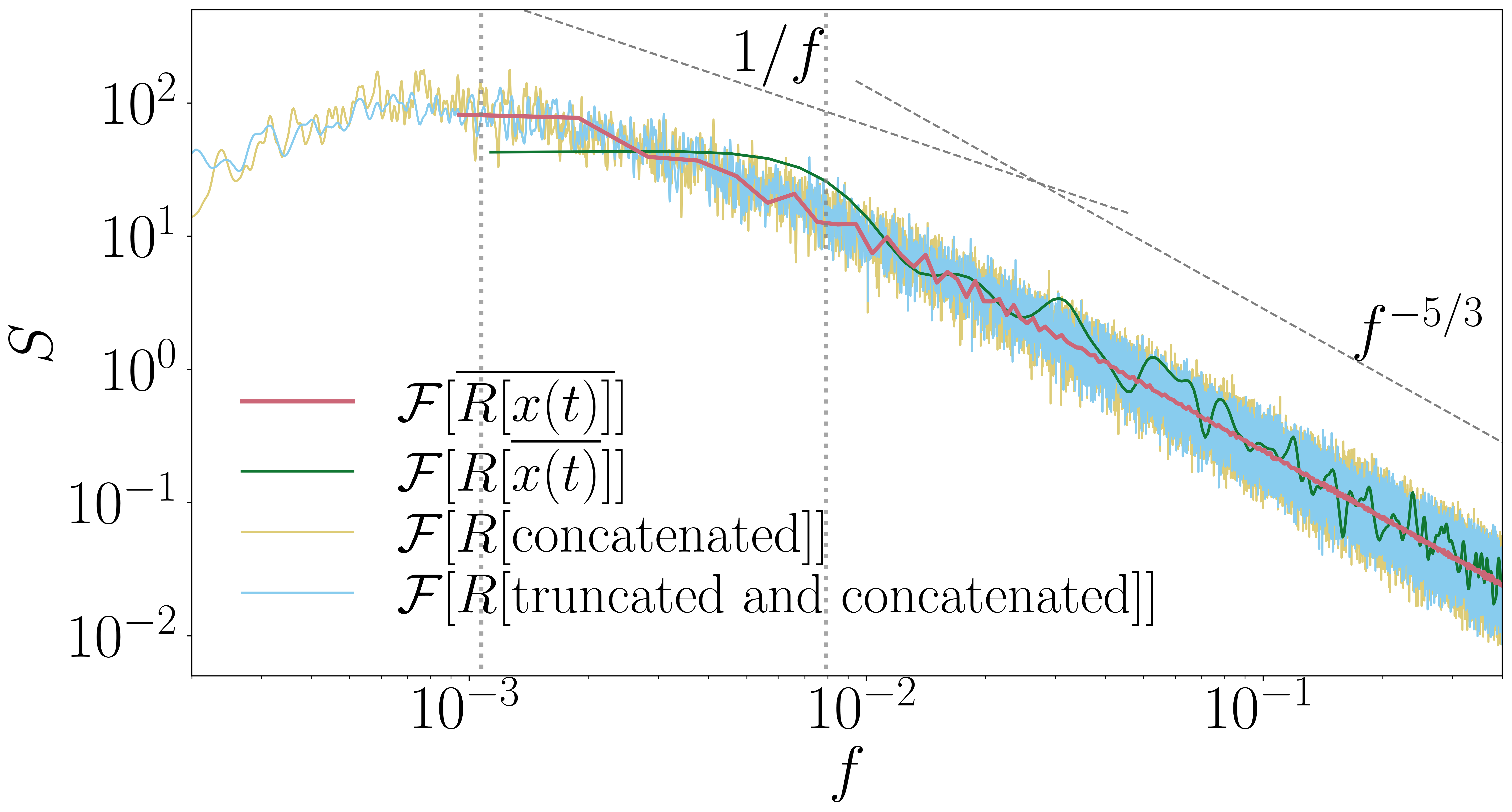}
    \includegraphics[angle=0,width=\columnwidth]{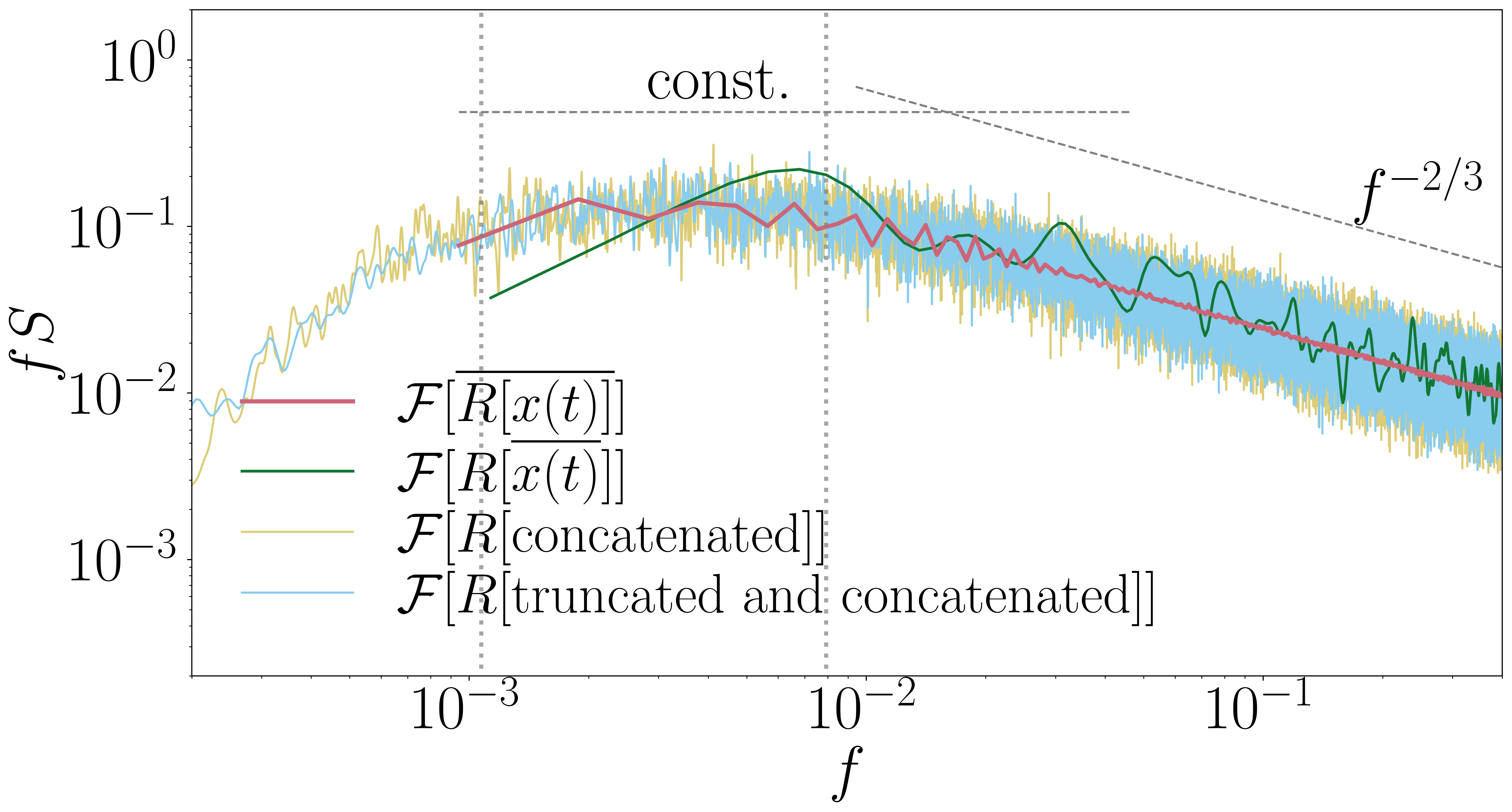}
    \caption{Top panel: power spectra resulting from lognormal correlation times with $\mu, \sigma^2 = 4, 1$, obtained through superposition methods 1 (in red), 2 (in green), 3 (in yellow), and 4 (in blue). Bottom panel: compensated spectra corresponding to those in the top panel. Dotted vertical lines mark $2\pi$-scaled boundaries of scale-invariant correlation time portion with $\theta = 0.5$ (see the text). Dashed grey lines are guiding power laws with index $-1$ and $-5/3$, respectively.}
\label{fig:Ssynthetic3}
\end{figure}

Figs.~\ref{fig:Ssynthetic2} and~\ref{fig:Ssynthetic3} show results for correlation time distributions of type (ii) with $\tau_1, \tau_2 = 20, 150$ and of type (iii) with $\mu, \sigma^2 = 4, 1$, respectively. These choices are intended to resemble the distributions observed in the 1 au solar wind, as will be shown in Fig.~\ref{fig:taudist} in Section~\ref{sec:observation}. Overall, the spectral behaviour is similar to that of the type (i) distribution case, with clear $1/f$ band appearing between the low-frequency flat band associated with uncorrelated fluctuations and the high-frequency $f^{-5/3}$ band. For the lognormal distribution case (Fig.~\ref{fig:Ssynthetic3}), the dotted vertical lines mark the boundaries $\exp(\mu \pm \sqrt{2\theta \sigma^2})$ of the scale-invariant region assuming a tolerable deviation factor of $\theta = 0.5$. As predicted, despite more gradual spectral transitions on both sides as compared to the type (ii) case, a distinct $1/f$ band still emerges between these boundaries for superposition Methods 1, 3, and 4.

\section{Superposition with 1 au Observation}
\label{sec:observation}

In the previous sections, we emphasized a generic mechanism for generating $1/f$ noise through an ensemble of signals. 
While not of the $1/f$-type in isolated samples, these signals collectively give rise to a $1/f$ spectrum. Assuming this procedure underlies the emergence of $1/f$ noise in the solar wind, a natural question is whether we can directly observe the process of $1/f$ formation at 1 au. As argued earlier, this is not feasible, as signals almost certainly already have undergone superposition of some form beginning at the base of the solar wind. However, by partitioning a long data record into subsets and comparing the resulting spectra from Methods 1, 2, and 3, we can assess how well the $1/f$ scaling is preserved (rather than created) under the superposition operation. Method 4 is redundant with respect to Method 3 in observational data, so we do not report its result.  

It should be made clear that the essential distinction between the synthetic time series in Section~\ref{sec:synthetic} and those measured at 1 au is that the latter may incorporate multiple correlated signals (thus time-scales) within a single interval. And there may be long-range effects such as the solar rotation that render correlations across separate intervals. It is thus impossible to isolate an ensemble of purely uncorrelated wind streams, which is only feasible in the realm of synthetic data. 
The superposition of a long-record low-frequency spectrum (Method 3) by averaging spectra from shorter subsets
of naturally occurring data may 
produce results that 
deviate in subtle ways from conclusions
based on our idealized model.

\subsection{Magnetic field data and analysis procedure}
\label{sec:observation_procedure}

The observational data set chosen as the testing ground for this study are magnetic field measurements at 1 au from ACE spacecraft~\citep{Stone98}. We analyse 12 yr of data spanning from 1998 February 5 to 2009 December 31, with records from midnight to midnight in Coordinated Universal Time, covering the minimum-to-minimum phase of the 23rd solar cycle~\citep{Hathaway15}. Since our goal is to probe the $1/f$ range at frequencies below the nominal inertial scales, the original 1-s resolution data collected by the Magnetometer~\citep{Smith98} aboard ACE are downsampled to 1-min cadence, with missing values flagged as `NAN'. No data gap interpolation is performed. The power spectrum of this 12-yr data set has previously been analysed in \cite{Wang26} and shows clear $1/f$ scaling in the trace from approximately $\unit[4 \times 10^7]{Hz}$ (around the solar rotation frequency) to approximately $\unit[2 \times 10^5]{Hz}$. We interpret this spectrum as reflecting superposition Method 3 and use it as a reference for comparison with spectra produced from Methods 1 and 2. 

For the analysis, we partition the data set independently into two subsets, one consisting of 1-d segments and the other of 10-d segments. 
A segment is excluded if more than 50 per cent of the measurements are missing. No distinction is made between segments dominated by fast or slow solar wind, nor are events such as interplanetary coronal mass ejections or heliospheric current sheet crossings removed. This is because the proposed reason of $1/f$ emergence involves the occurrences of rare events (at the tail of the scale-invariant or lognormal distributions) -- or mixtures of them -- each contributing correlated signals. This results in a total of 4259 1-d intervals and 426 10-d intervals. The former set is used to estimate the interplanetary turbulence correlation time distribution, and both sets are used to reconstruct the 12-yr spectrum. 

For Method 1, component-wise magnetic fluctuation autocorrelations for each interval are first computed as in the numerator of Eq.~\ref{eq:R} through the Blackman-Tukey algorithm~\citep{Blackman58, Matthaeus82-invariants}, which are then summed to get the trace. As in the test with synthetic data, $j$ indexes temporal lag $\tau_j = j \Delta t$, $\Delta t = \unit[60]{s}$, and ranges from 0 to 30 per cent of the segment duration. The magnetic spectrum is computed from the correlation through the same procedure adopted for the synthetic data, following Eq.~\ref{eq:psd_dsc} and then cosine-tapering and zero-padding. In the next subsection, we compare the ``averaged-then-normalized'' trace spectrum with the ``normalized-then-averaged'' one to see whether $1/f$ emerging from superposition withstands variation in fluctuation variance. We repeat the analysis for 50 randomly selected 10-d intervals separated in time, thereby reducing cross-interval correlations.

For Method 2, the magnetic field measurements are averaged across all 10-d intervals to produce a single representative 10-d time series. The correlation trace of the fluctuation and the corresponding spectrum are then computed from this averaged data set.

The turbulence correlation time is determined from the 1-d intervals as the $e$-folding time of the correlation trace $R$, i.e., the time lag where the correlation decreases by a factor of $1/e$. The correlation trace is first smoothed 3 times through the moving-average filter (as described in Section~\ref{sec:synthetic_analysis}) to mitigate fluctuations. If the correlation does not reach the $1/e$ level within the first 90 per cent of its domain, we perform linear fit to $\ln{R}$ versus $\tau$ for the first 50 per cent of the correlation to acquire the value of $\tau$ where $\ln{R}/\ln{R(0)} = -1$, similar to the approach in \citet{Isaacs15}.

\subsection{Results}

\begin{figure}
\centering
    \includegraphics[angle=0,width=\columnwidth]{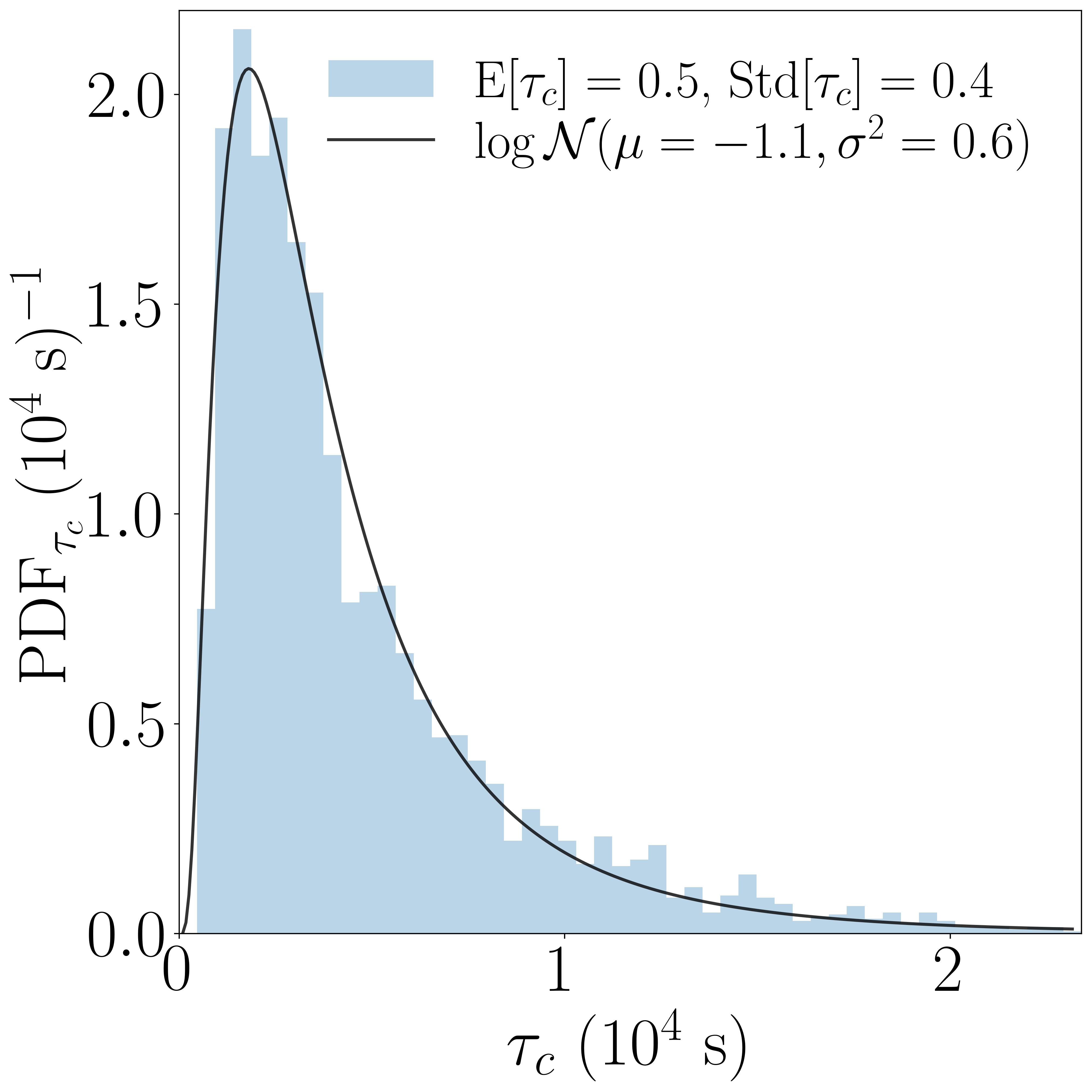}
    \caption{Distribution of magnetic field correlation time evaluated from 24-h intervals. Ensemble expectation value (E) and standard deviation (Std) are listed in legend. Solid curve represents best-fitting lognormal distribution with parameters also reported in legend.}
\label{fig:taudist}
\end{figure}

We show in Fig.~\ref{fig:taudist} the distribution of magnetic field correlation times at 1 au. The correlation times follow a lognormal distribution, consistent with previous findings~\citep{Ruiz14, Isaacs15, cuesta2022isotropization}. Moreover, \citet{Isaacs15} reported that both the mean and variance of this distribution increase with the duration of the averaging interval, with the trend qualitatively leading to the distribution shown in Fig.~\ref{fig:taudist}, obtained using 24-h intervals. We emphasize that the $1/e$ method employed here yields the local turbulence correlation time. Correlations on day-long and longer time-scales, which may also contribute to the formation of $1/f$ noise, are not represented in Fig.~\ref{fig:taudist}.

\begin{figure}
\centering
    \includegraphics[angle=0,width=\columnwidth]{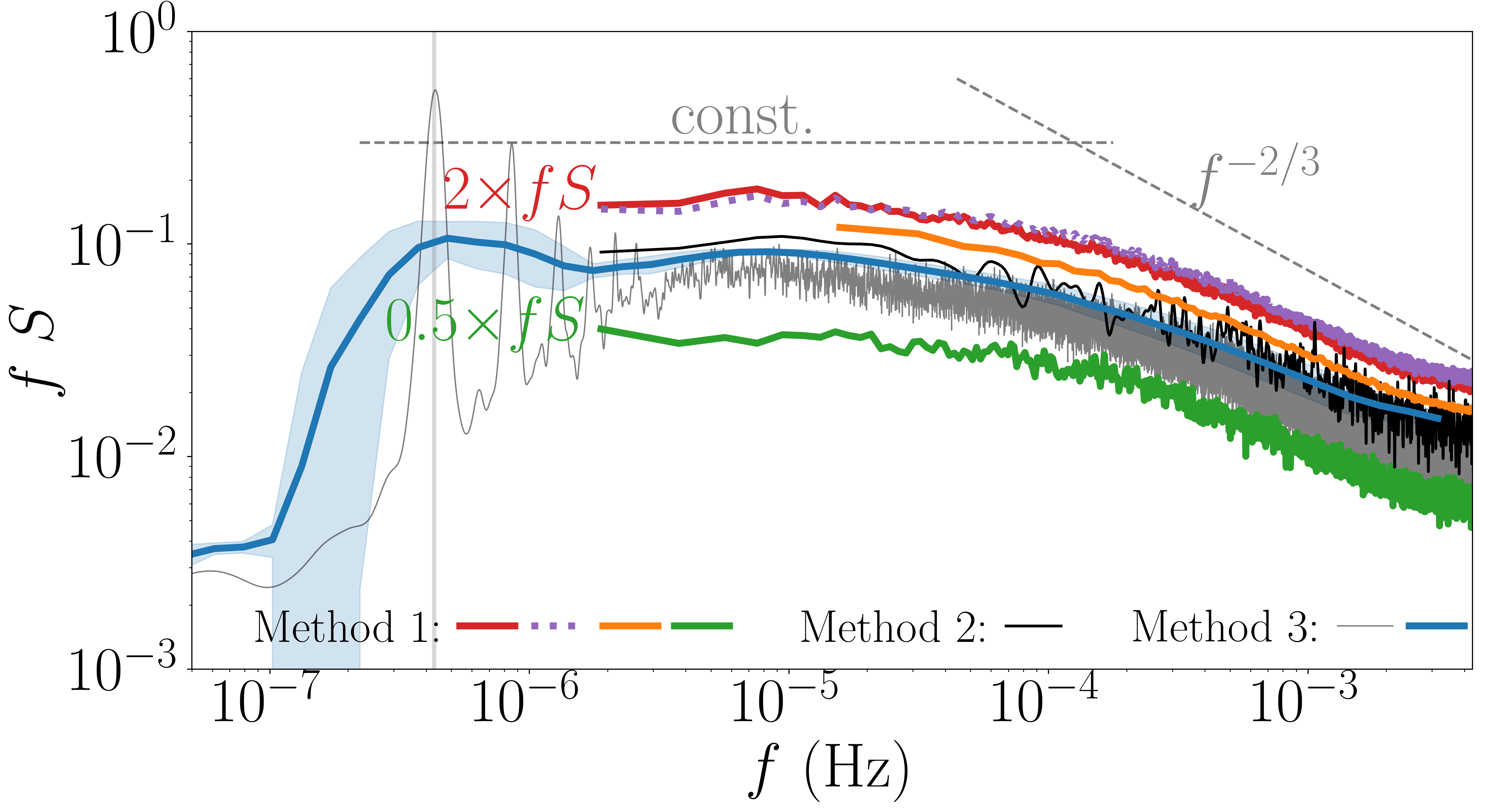}
    \caption{Recovery of 1/f with different superposition methods. Method 1: Averaged-then-normalized spectrum from 10-d intervals (red); normalized-then-averaged spectrum (dotted purple); averaged-then-normalized spectrum from randomly selected 10-d intervals (green); averaged-then-normalized spectrum from 1-d intervals (orange).
    Method 2: Spectrum from averaged 10-d intervals (black).
    Method 3: Spectrum from unsegmented 12-yr data (grey); integrated spectrum and associated uncertainty from 12-yr data~\citep[blue curve and shaded area, respectively, from][]{Wang26}. Dashed lines indicate constant and $f^{-2/3}$ power laws. Vertical line indicates 27-d solar rotation frequency.}
\label{fig:Sace}
\end{figure}

Fig.~\ref{fig:Sace} shows the frequency-compensated spectra $fS$ obtained from superposition Methods 1, 2, and 3 using ACE data. For completeness, we present the spectrum from the unsegmented 12-yr time series (Method 3), as published in \citet{Wang26}, 
alongside our new results: the raw spectrum is smoothed 50 times and shown in grey; the frequency-integrated spectrum and its associated uncertainty are shown in blue. We show four spectra derived from averaging correlations (Method 1): the averaged-then-normalized (normalized-then-averaged) spectrum from the full set of 10-d intervals is shown in red (purple), multiplied by a factor of 2 for visualization; the averaged-then-normalized spectrum from randomly selected 10-d intervals is shown in green, multiplied by a factor of 0.5; the averaged-then-normalized spectrum from the full set of 1-d intervals is smoothed one time to suppress low-frequency fluctuations and is shown in orange. The spectrum from averaged 10-d time series (Method 2) is smoothed 10 times and shown in black. 

All spectra in Fig.~\ref{fig:Sace} exhibit broadly similar shapes across their overlapping frequency domains and consistently display clear $1/f$ behaviour. The presence of the Method 2 spectrum (black) within this consensus -- unlike in the synthetic tests -- likely reflects the relatively negligible cross-interval correlations compared to the autocorrelations within individual 10-d intervals. This 
is in accord with the observation that solar wind magnetic field correlations rarely return to the $1/e$ level after de-correlating on several-hour time-scales, even accounting for the moderate correlation increase at time-scales associated with the solar rotation~\citep{Wang26}.

The averaged-then-normalized spectrum (red) closely matches the normalized-then-averaged spectrum (purple), indicating that the emergence of a $1/f$ band is not sensitive to the variation in fluctuation power $\langle x^2 \rangle-\langle x \rangle^2$ in the 10-d data set, even though this quantity is known to follow lognormal distributions~\citep{Ruiz14, Isaacs15, Sharma23, Chhiber25_vKH}.

\section{Conclusion and Discussion}
\label{sec:conclusion}

\textmod{Acknowledging the limitations of the generic superposition principle as developed in \citet{Machlup81} -- namely, its reliance on correlation-domain operations -- we examine the feasibility of superposition carried out directly in the time domain. To this end, we test several implementations using synthetic time series constructed with scale-invariant properties. Time-domain superposition is more plausibly associated with physical processes occurring in natural systems such as the solar coronal wind and the solar interior dynamo.}

\textmod{Based on this framework}, we report the generation and preservation of $1/f$ signals through four types of superposition operations: (1) averaging correlation functions, (2) averaging time series, (3) concatenating time series, and (4) concatenating randomly truncated time series. \textmod{We use synthetic data to demonstrate the feasibility of generation, and solar wind measurements to demonstrate the preservation.} Methods 1 and 3 display very similar spectral shape in the synthetic and the measured 1 au data, respectively, which is expected as data elongation translates directly to averaging the autocorrelations within the shared frequency domain. Spectral similarity between Methods 3 and 4 may reflect the fact that the correlation time is characteristic to an interval and is barely affected by interval truncation. 

These results are \textmod{broadly} relevant to spectral analysis of observational data, as long-duration heliospheric measurements inherently consist of a mixture of signals with \textmod{extensively} distributed correlation scales, and are \textmod{frequently} subject to truncation or concatenation during data collection and analysis.

\textmod{We note that the synthetic data construction is highly idealized. In particular, the very low-frequency spectral density is flat, corresponding to uncorrelated signals at those time-scales. This to some degree differs from solar wind observations, where correlations persist for much longer times. It is noteworthy that recent analyses~\citep{Davis23, Huang25} of solar wind data indicate that spectral behaviour just below the break frequency exhibits variability and does not always conform precisely to $1/f$. Our idealized construction therefore represents an extreme limiting case in which none of the individual samples exhibits $1/f$ scaling. While this differs from the actual solar wind measurements, it is motivated by the desire to remain close to the analytical model.}

\textmod{Any sub-coronal superposition process is not directly observable in situ, so at best we assess whether $1/f$ signals, once established, can be preserved under subsequent superposition operations using local measurements. Possible generation in the corona~\citep[as suggested in, e.g., ][]{Matthaeus86, Mullan90} has been argued to depend on scale-invariant reconnection processes. It is also conceivable that time-domain signals are superposed within the solar interior, as suggested by magnetogram $1/f$ observations in \citet{Nakagawa74}.}

The observed $1/f$ band in Fig.~\ref{fig:Sace} approximately overlaps with the range expected from the local turbulence correlation times in Fig.~\ref{fig:taudist}. But it extends to frequencies as low as $\unit[10^{-6}]{Hz}$, comparable to the superharmonics of the solar rotation period. This indicates that long-range correlations on day- to month-time-scales play a significant role in shaping the $1/f$ spectrum. Such long correlation times cannot be captured from the conventionally used $1/e$ method, nor can they be estimated via correlation integration~\citep{Matthaeus05}, since the trace of the magnetic correlation oscillates around zero with solar-rotation periodicities~\citep{Matthaeus82-convergence, Wang26} and is therefore not integrable. We defer a detailed treatment of long-range scales to future work.

Superposition mechanisms of the type 
covered in the above sections are unlikely to operate locally in the 
super-Alfv\'enic wind to produce the observed interplanetary 
$1/f$ signal. This assertion is based on the 
realization that 
any proposed local $1/f$ generation mechanism faces causality and range-of-influence issues, and is not able to explain $1/f$ behaviour down to the lowest observed $\unit[10^{-6}]{Hz}$ frequency range~\citep[see e.g.][]{Zhou90, Chhiber2018thesis, Wang24_1overf}. 
Conversely, 
presently available
evidence points to 
a source of the
constituents of the $1/f$
magnetic field signal that 
lies deep within the corona or beneath the photosphere~\citep{Matthaeus86, Zhou90, Wang26}.
Theoretical attention then necessarily turns to identification 
of the responsible dynamical mechanisms.
While specific inquiries along these
lines are deferred to future studies, 
it seems likely that 
explanations might be developed based on dynamo theory~\citep{Moffatt69}, 
self-organized criticality~\citep{Bak88}, and inverse cascade~\citep{Frisch75}.
Indeed, connections between these frameworks and $1/f$ noise are already present in the
literature~\citep{Ponty04, Dmitruk07, Dmitruk11, Dmitruk14, Lamarre25}.

Exploring $1/f$ magnetic field fluctuation in the context of these frameworks is an important direction for future study.
The ultra-low frequency variations that comprise the $1/f$ signal 
provide a physical context and background in which 
extreme interplanetary events
and space weather occur, thus
providing substantial 
motivation.
Future investigations may benefit from appeal to the superposition principles set out in the present series of exercises.
In particular, it may be possible to characterize 
the relevant superposed real-space structures 
by analysis of imaging data from the PUNCH mission~\citep{Deforest22}, or based on statistical analysis of multi-spacecraft data from HelioSwarm~\citep{Klein23} or the L1 constellation of spacecraft, including the IMAP mission, near 1 au~\citep{McComas25}.


\section*{Acknowledgements}

This research was partially supported by
the U.S. National Science Foundation, award PHY-2108834, through the NSF/DOE Partnership in Basic Plasma Science and Engineering, 
by NASA LWS grants 80NSSC20K0377 (subcontract 655-001) and 80NSSC22K1020, by the NASA IMAP project at UD under subcontract SUB0000317 from Princeton University, by the NASA/SWRI PUNCH subcontract N99054DS at the University of Delaware, and by the NASA HSR grant 80NSSC18K1648.

\section*{Data Availability}

All study data are included in the article. There are no new observational data generated or analysed in support of this research. 

\bibliographystyle{mnras}


\appendix
\section{Generation of synthetic data}
\label{app}

We first give a basic overview of our discrete Fourier transform and power spectral density formalism. Suppose the discrete, bounded time series we wish to generate is represented by $x_j$ on the time grid $t_j = j \Delta t$, $j = 0, \cdots, N-1$, over the time domain $[0, T]$ where $T = N \Delta t$. Let $X_m$ be the discrete Fourier transform of $x_j$ on the frequencies $f_m$, so
\begin{equation}
    X_m = \Delta t \sum_{j=0}^{N-1} e^{-i2\pi m j/N} x_j,
\label{eq:dft}
\end{equation}
\begin{equation}
    f_m = \frac{m}{N\Delta t}, \indent m = -\frac{N}{2}+1, \cdots, \frac{N}{2}.
\end{equation}
The power spectral density $S_m$ is related to $X_m$ by $S_m = \langle X_m^* X_m \rangle/T$ under the constraint that $\tau_c \ll T$, where the asterisk stands for taking the complex conjugate. By Hermitian symmetry, $X_m = X^*_{-m}$.

We now construct the discrete, normalized power spectrum:
\begin{equation}
    S_m = \frac{1}{C} 
    \begin{cases}
        0 & f_m = 0\\
        1 & 0 < |f_m| < f_\mathrm{break}\\
        \left( \frac{f_m}{f_\mathrm{break}} \right)^{-\alpha} & |f_m| \geq f_\mathrm{break}
    \end{cases}
\label{eq:Sm}
\end{equation}
where
\begin{equation}
    C = \frac{1}{N \Delta t} \left[ \left(\frac{1}{2 \Delta t f_\mathrm{break}} \right)^{-\alpha} + 2 \lfloor N \Delta t f_\mathrm{break} \rfloor + 2 \sum_{m=\lfloor N \Delta t f_\mathrm{break} +1 \rfloor}^{N/2-1} \left( \frac{m}{N \Delta t f_\mathrm{break}} \right)^{-\alpha} \right],
\label{eq:normC}
\end{equation}
where $-\alpha < -1$ is the power-law index. Note that the constraint of $S_0 = 0$ ensures $\langle x_j \rangle = 0$, and the normalization ensures $\langle x_j^2 \rangle = 1$. Note also that when calculating the normalization constant, an integration of equation \ref{eq:Sm} instead of a summation as in equation \ref{eq:normC} is feasible, where the result is $C = 2 f_\text{break} \left[\alpha/\left(\alpha + 1\right)\right]$. But this leads to numerical errors due to the discreteness of the spectrum.

Now, $S_m$ is real by definition while $X_m$ may not be, and $|X_m| = (T S_m)^{1/2}$. Thus we assign random phase $\phi_m$ to each $|X_m|$, with $\phi_{N/2} \equiv 1$ and $\phi_m \in [0, 1]$ uniformly distributed elsewhere. The first condition ensures that $X_{N/2}$, the amplitude at the Nyquist frequency, must be real. So $X_m$ takes the form
\begin{equation}
    X_m = \left( \frac{T}{C} \right)^{1/2}
    \begin{cases}
        0 & f_m = 0\\
        e^{i 2\pi \phi_m} & 0 < f_m < f_\text{break}\\
        e^{i 2\pi \phi_m} \left( \frac{f_m}{f_\text{break}} \right)^{-\alpha/2} & f_m \geq f_\text{break}
    \end{cases}
\end{equation}
for the positive frequencies with $m = 0, \cdots, N/2$, and $X_{-m} = X_m^*$ for the negative frequencies with $m = -N/2+1, \cdots, -1$.

Finally, the random time series $x_j$ can be constructed by inverse discrete Fourier transforming $X_m$:
\begin{equation}
    x_j = \frac{1}{N \Delta t} \sum_{m = -N/2+1}^{N/2} e^{i 2\pi m j/N} X_m,
\label{eq:idft}
\end{equation}
\begin{equation}
    t_j = j \Delta t, \indent j = 0, 1, \cdots, N-1.
\end{equation}
Eqs.~\ref{eq:dft} and~\ref{eq:idft} apply to any Fourier transform pairs.

\bsp	
\label{lastpage}

\end{document}